\documentclass[reprint,amsmath,amssymb,aps,prl,floatfix,superscriptaddress]{revtex4-1}
\usepackage{graphicx}
\usepackage{dcolumn}
\usepackage{bm}
\usepackage[colorlinks,breaklinks,citecolor=blue,linkcolor=blue,urlcolor=black]{hyperref}

\begin{document}

\title{Quantum Zeno effects across a parity-time symmetry breaking transition in atomic momentum space}%

\author{Tao Chen}
\author{Wei Gou}
\author{Dizhou Xie}
\author{Teng Xiao}
\affiliation{%
Interdisciplinary Center of Quantum Information, State Key Laboratory of Modern Optical Instrumentation, and Zhejiang Province Key Laboratory of Quantum Technology and Device of Physics Department, Zhejiang University, Hangzhou 310027, China
}%
\author{Wei Yi}
\affiliation{CAS Key Laboratory of Quantum Information, University of Science and Technology of China, Hefei 230026, China}
\affiliation{CAS Center For Excellence in Quantum Information and Quantum Physics, Hefei 230026, China}
\author{Jun Jing}
\email{jingjun@zju.edu.cn}
\affiliation{%
Department of Physics, Zhejiang University, Hangzhou 310027, China
}%
\author{Bo Yan}
\email{yanbohang@zju.edu.cn}
\affiliation{%
Interdisciplinary Center of Quantum Information, State Key Laboratory of Modern Optical Instrumentation, and Zhejiang Province Key Laboratory of Quantum Technology and Device of Physics Department, Zhejiang University, Hangzhou 310027, China
}%
\affiliation{%
 Collaborative Innovation Centre of Advanced Microstructures, Nanjing University, Nanjing, 210093, China
}%
\affiliation{%
 Key Laboratory of Quantum Optics, Chinese Academy of Sciences, Shanghai, 200800, China
}

\date{\today}

\begin{abstract}
We experimentally study quantum Zeno effects in a parity-time (PT) symmetric cold atom gas periodically coupled to a reservoir.
Based on the state-of-the-art control of inter-site couplings of atoms in a momentum lattice, we implement a synthetic two-level system with passive PT symmetry over two lattice sites, where an effective dissipation is introduced through repeated couplings to the rest of the lattice. Quantum Zeno (anti-Zeno) effects manifest in our experiment as the overall dissipation of the two-level system becoming suppressed (enhanced) with increasing coupling intensity or frequency. We demonstrate that quantum Zeno regimes exist in the broken PT symmetry phase, and are bounded by exceptional points separating the PT symmetric and PT broken phases, as well as by a discrete set of critical coupling frequencies.
Our experiment establishes the connection between PT-symmetry-breaking transitions and quantum Zeno effects, and is extendable to higher dimensions or to interacting regimes, thanks to the flexible control with atoms in a momentum lattice.
\end{abstract}

\maketitle

The decay of an unstable quantum system can be suppressed by frequent projective measurements, whose back actions repeatedly interrupt
the time evolution of the system.
Such a phenomenon, famed as the quantum Zeno effect, has been experimentally observed in various physical systems~\cite{Itano1990,Fischer2001,Streed2006,Kilina2013,Harrington2017}, and has found widespread utilities in quantum information~\cite{Wang2008,Maniscalco2008,Shao2009,Chand2010,deLange2010,Smerzi2012,Silva2012,Zhu2014,Kalb2016,Hacohen2018,Szombati2020} and quantum simulation~\cite{Schafer2014,Signoles2014,Bretheau2015,Barontini2015,Do2019}.
In a complementary fashion, with an appropriate repetition rate of measurements, the decay of the system can also be enhanced under what is known as the anti-Zeno effect~\cite{Kofman2000}.
Intriguingly, both quantum Zeno and anti-Zeno effects are alternatively accessible through continuous strong couplings or fast unitary kicks~\cite{Facchi2008, Schafer2014, Streed2006} that couple a system to an auxiliary Hilbert space. With the auxiliary Hilbert space playing the role of environment, these processes give rise to dissipative system-reservoir couplings, under which the time evolution of the system is effectively driven by a non-Hermitian Hamiltonian.

Although evidence of quantum Zeno effects have been theoretically demonstrated and experimentally observed in non-Hermitian settings~\cite{Cui2019,Gou2020},
surprisingly little is discussed on its interplay with parity-time (PT) symmetry, despite the latter being a ubiquitous property of non-Hermitian systems while holding great promise for future applications \cite{Bender1998,Konotop2016,Ganainy2018}. A PT symmetric, non-Hermitian system possesses two distinct phases: the parity-time symmetric (PTS) phase, with entirely real eigenenergy spectrum; and the parity-time broken (PTB) phase, where eigenenergies are complex in general. The two phases are separated by exceptional points, with coalescing eigenstates and eigenenergies.
While quantum Zeno effects naturally emerge in the deep PTB regime which can be mapped to an open system possessing continuous and strong coupling with a dissipative reservoir~\cite{Naghiloo2020}, the fate of quantum Zeno (anti-Zeno) effects is less well-known in the PTS regime or near exceptional points, both of which typically occur at much smaller dissipation strengths~\cite{Wu2019, Li2019, Naghiloo2020}.
A very recent theoretical study shows that exceptional points of a PT symmetric Hamiltonian also mark the boundary between quantum Zeno and anti-Zeno regimes~\cite{Li2020}, suggesting a deep connection between the two previously independent fields of study.
Here we experimentally confirm such a connection in a PT symmetric, synthetic two-level system, embedded in a momentum lattice of cold atoms.

\begin{figure}[]
\includegraphics[width=0.5\textwidth]{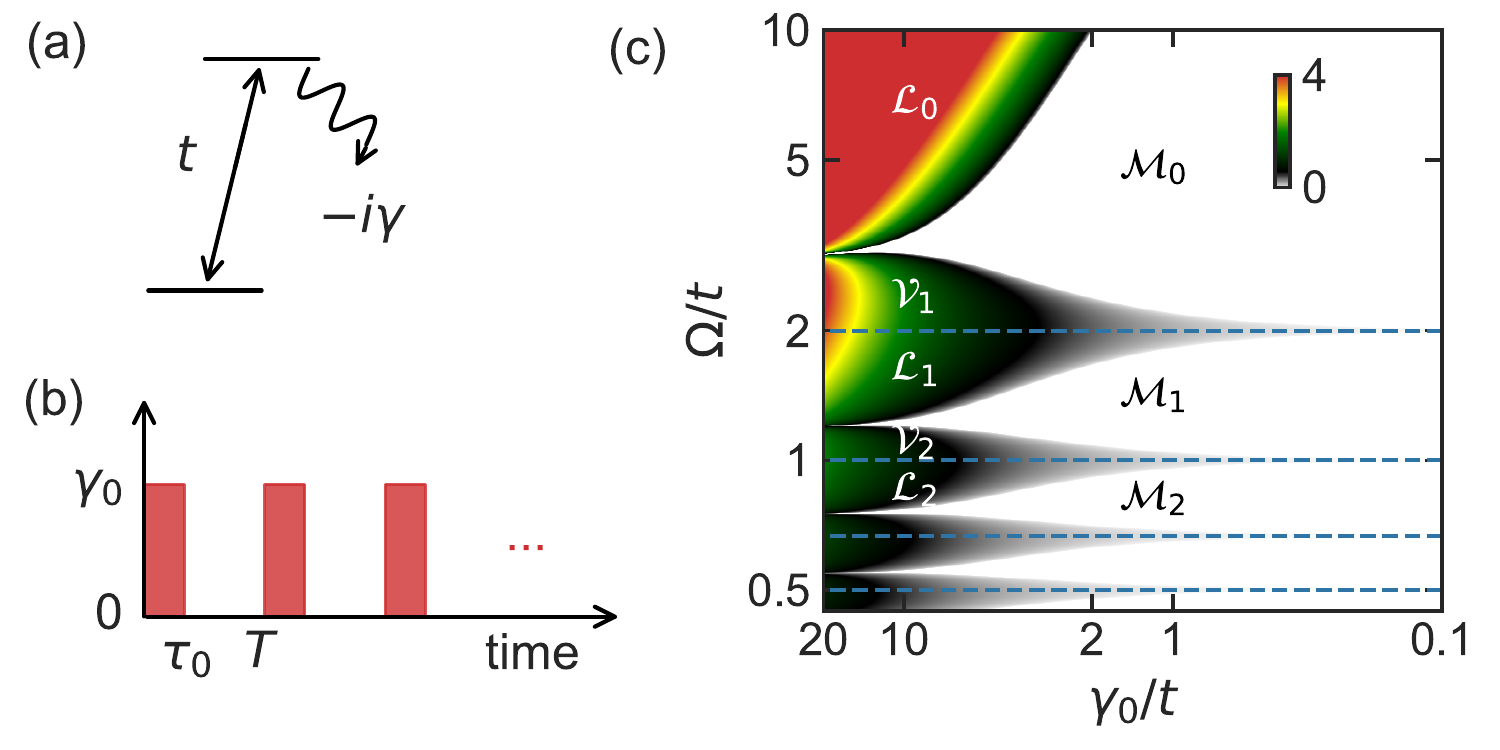}
\caption{\label{fig1} (Color Online)
(a) Schematic illustration of the dissipative two-level system under Hamiltonian (\ref{eq1}).
(b) The loss rate $\gamma(\tau)$ in (a) is time-periodic with a square-wave modulation. The modulation period $T=2\pi/\Omega$ ($\Omega$ is the modulation frequency), and $\gamma(\tau)=\gamma_0$ during the duty time interval $[0, \tau_0]$.
(c) Theoretical phase diagram in the $\Omega$--$\gamma_0$ plane. Color contour shows the
dimensionless parameter $\lambda$ (see main text for definition). Here we set $\tau_0 t/(2\pi)=0.1$. Regions with vanishing $\lambda$ correspond to the PTS phase ($\mathcal{M}_j$), while the colored regions ($\mathcal{L}_j$ and $\mathcal{V}_j$) correspond to the PTB phase. The horizontal dashed lines indicate critical coupling frequencies separating the $\mathcal{L}_j$ (quantum Zeno) and $\mathcal{V}_j$ (quantum anti-Zeno) regimes.
}
\end{figure}

We focus on a two-level system under time-periodic dissipation~\cite{Li2020}, as illustrated in Fig.~\ref{fig1}(a). The time-dependent Hamiltonian is
\begin{equation}\label{eq1}
 H/\hbar = -\frac{i\gamma(\tau)}{2}\mathbb{I} + t\sigma_{\rm x} + \frac{i\gamma(\tau)}{2}\sigma_{\rm z},
\end{equation}
where $\mathbb{I}$ and $\sigma_{x,z}$ are the identity and Pauli matrices respectively, $\tau$ is the evolution time, and $t$ is the inter-state coupling rate.
The time-periodic dissipation rate $\gamma$ is given by
\begin{equation}\label{eq2}
 \gamma(\tau) = \begin{cases}
	\gamma_0 & jT \le \tau < jT+\tau_0 \\
	0 & jT +\tau_0 \le \tau < (j+1)T
 \end{cases} \ \  ,
\end{equation}
where $j\in\mathbb{Z}$, the modulation period $T=2\pi/\Omega$ with $\Omega$ the modulation frequency, $\gamma_0$ characterizes the modulation intensity, and $\tau_0$ is the duty time interval with nonzero $\gamma$ in each cycle; see Fig.~\ref{fig1}(b).

Hamiltonian (\ref{eq1}) is passive PT symmetric, in that it is purely dissipative, but is directly related to the standard PT symmetric Hamiltonian $H_{PT}=t\sigma_{\rm x} + \frac{i\gamma}{2}\sigma_{\rm z}$ with balanced gain and loss. Explicitly, $\mathcal{PT}H_{PT}\mathcal{PT}^{-1}=H_{PT}$, with the PT symmetry operator $\mathcal{PT}=\sigma_x  \mathcal{K}$ where $\mathcal{K}$ is complex conjugation.
Since PT symmetry of $H$ is determined by the imaginary parts of the quasienergies $\hbar\epsilon_\pm$ of the corresponding Floquet Hamiltonian~\cite{Li2019, Li2020},
we adopt a dimensionless parameter $\lambda = |{\rm Im}(\epsilon_+ - \epsilon_-)|/t$ to characterize the PT-symmetry breaking transition.
Here $e^{-i\epsilon_{\pm}T/\hbar}$ are eigenvalues of the non-unitary time-evolution operator $U=\hat{\mathcal{T}}e^{-i\int_0^T H(\tau)/\hbar d\tau}$,
where $\hat{\mathcal{T}}$ is the time-ordering operator.
For $\lambda=0$, the system lies in the PTS phase, while $\lambda>0$ corresponds to a PTB phase.

\begin{figure}[tbp]
\includegraphics[width=0.5\textwidth]{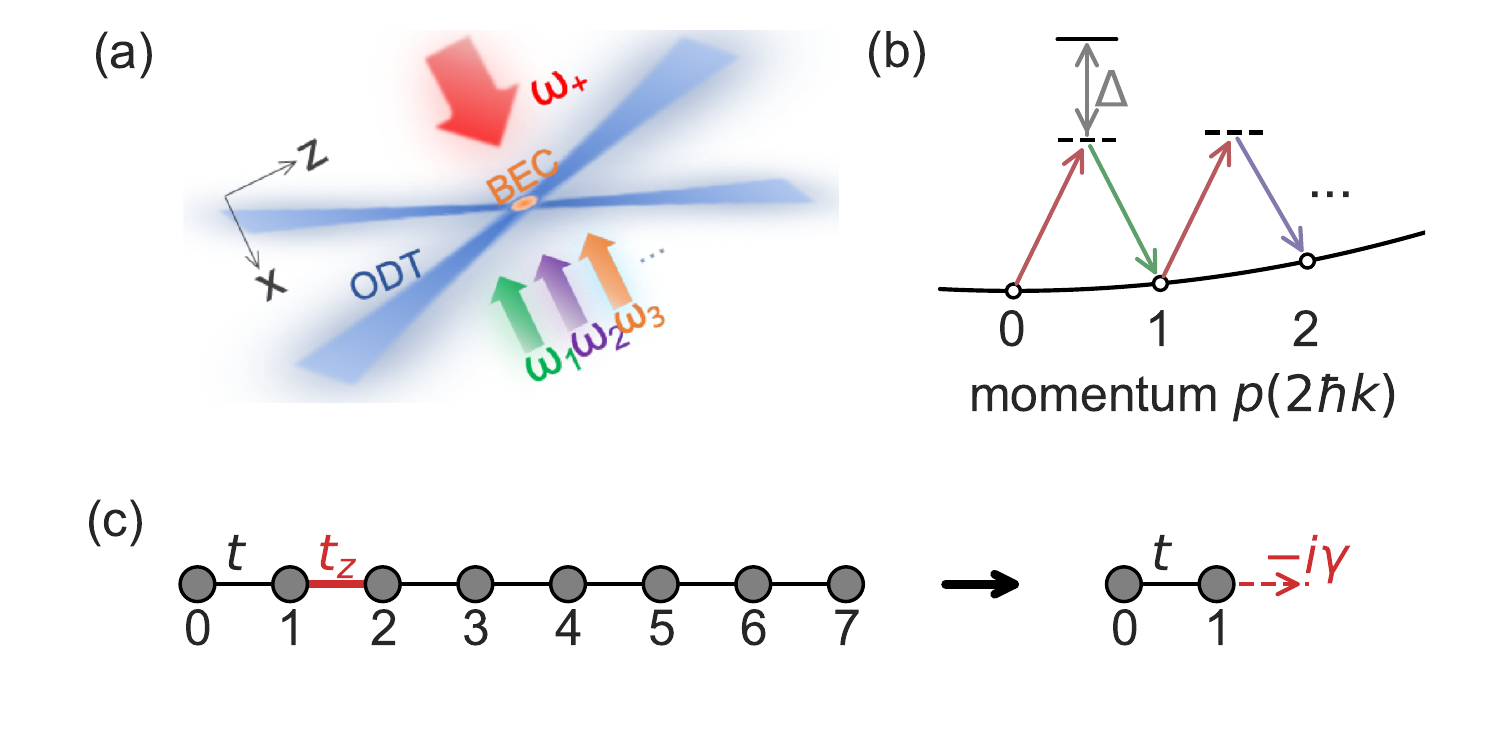}
\caption{\label{fig2} (Color Online)
(a) Schematic of the experimental setup. A Bose-Einstein condensate interacts with a pair of counter-propagating Bragg lasers in an optical dipole trap~\cite{suppinfo}. (b) Each Bragg laser pair triggers a resonant two-photon Bragg transition, coupling two neighbouring momentum states along the momentum lattice.
(c) The resulting $8$-site momentum lattice is mapped into a dissipative two-level system with tunable loss rate $\gamma(\tau)$, by treating the sites $\{|0\rangle,|1\rangle\}$ as the system, and $|n\ge 2\rangle$ as a reservoir.
}
\end{figure}

Figure~\ref{fig1}(c) shows a numerically calculated phase diagram with a fixed $\tau_0 t$. The white PTS region is separated into several blocks (marked as $\mathcal{M}_j$), by a series of critical modulation frequencies $\Omega_j = 2t/j$ ($j\in\mathbb{N}^+$) at which the PTS phase vanishes and PT-symmetry breaking is at its maximum.
The colored PTB regimes are further divided by the critical modulation frequencies into $\mathcal{L}_j$ and $\mathcal{V}_j$ regions, respectively corresponding to quantum Zeno and anti-Zeno regimes, as we explicitly demonstrate later.
For any fixed $\Omega \neq \Omega_j$, a PTS to PTB transition ($\mathcal{M}_j\to\mathcal{L}_j$ or $\mathcal{M}_j\to\mathcal{V}_{j+1}$) is crossed when increasing $\gamma_0$ from weak to strong. However, for a fixed $\gamma_0$, the PTS and PTB phases alternate ($\mathcal{M}_j\to\mathcal{L}_j\to\mathcal{V}_j\to\mathcal{M}_{j-1}\to\mathcal{L}_{j-1}\to\cdots$) with increasing modulation frequency $\Omega$.
While the phase diagram is distinct from that of PT symmetric systems under a continuous dissipation~\cite{Naghiloo2020}, a crucial observation is that quantum Zeno regimes exist in the PTB phases, and are bounded by critical frequencies, as well as by exceptional points pertaining to the PT symmetry breaking transitions.

\begin{figure*}[]
\includegraphics[width=\textwidth]{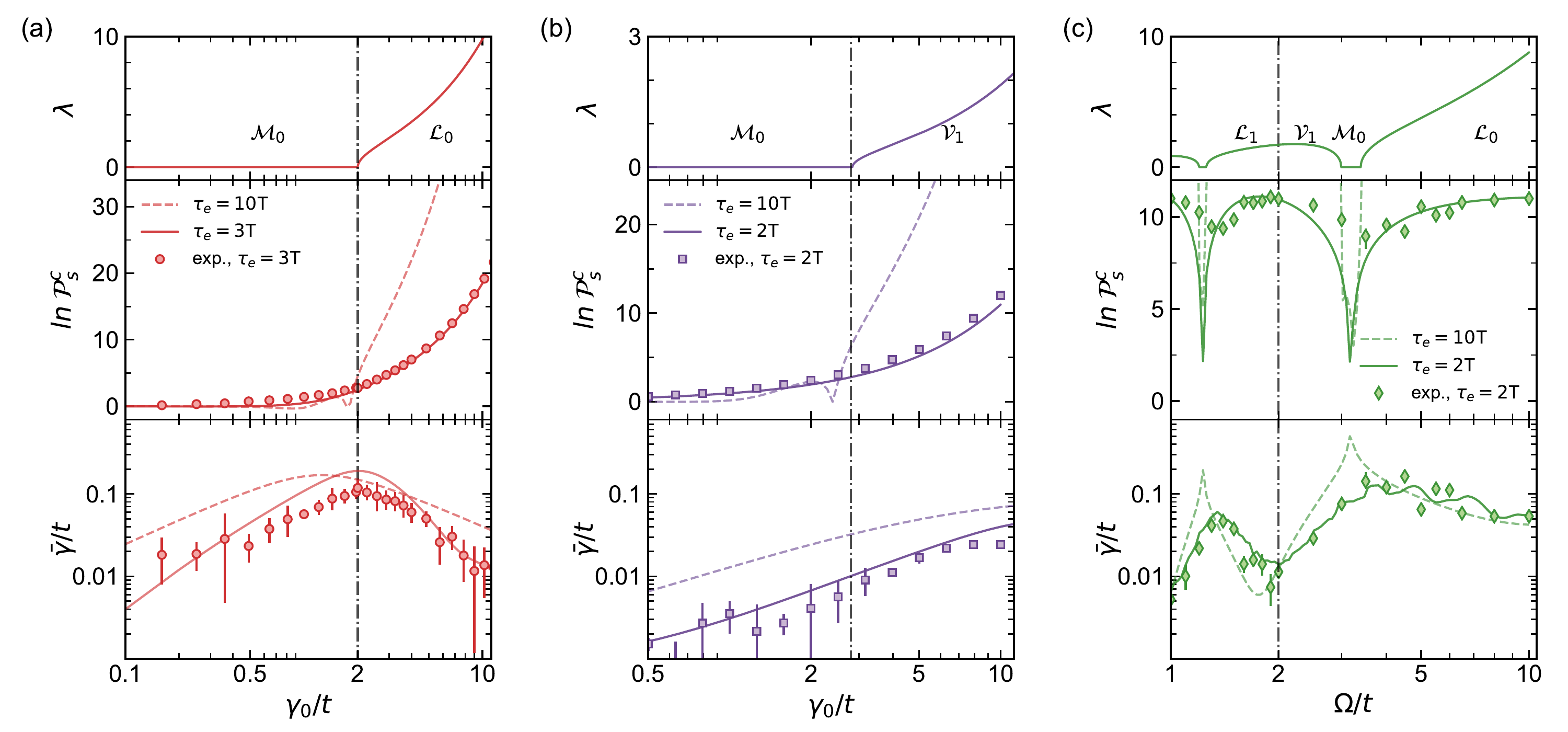}
\caption{\label{fig3} (Color Online)
Observation of the correspondence between quantum (anti-)Zeno effects and the PT phases by varying the kick parameters, with numerically calculated dimensionless parameter $\lambda$ (upper panel), measured corrected probability $\mathcal{P}_s^c$ (middle panel), and the measured averaged loss rate $\bar{\gamma}$ (lower panel).
(a) Dependence of $\lambda$, $\mathcal{P}_s^c$, and $\bar{\gamma}$ on the kick intensity $\gamma_0$ under a large kick frequency $\Omega/t=10$.
The exceptional point is at $\gamma_0/t=2$ (dash-dotted line).
(b) Dependence of $\lambda$, $\mathcal{P}_s^c$, and $\bar{\gamma}$ on $\gamma_0$ during the duty time under a small kick frequency $\Omega/t=2.5$.
The exceptional point is at $\gamma_0/t=2.8$ (dash-dotted line).
(c) Dependence of $\lambda$, $\mathcal{P}_s^c$, and $\bar{\gamma}$ on the kick frequency $\Omega$ with a fixed $\gamma_0/t \sim 9$.
The evolution time $\tau_e=3T$ for (a), while $\tau_e=2T$ for (b) and (c).
For all experiments, we take $t=2\pi\times 1.03(2)~{\rm kHz}$, and the duty time is $\tau_0 t/(2\pi)=0.1$.
The solid (dashed) lines in the middle panels are numerical simulations with respective experiment-used $\tau_e$ (longer $\tau_e=10T$), while the solid (dashed) lines in the lower panels are numerical simulations with Hamiltonian (\ref{eq1}) (an effective Hamiltonian including higher-order, non-resonant coupling terms~\cite{suppinfo}).
The vertical dash-dotted line in (c) indicates the critical kick frequency $\Omega_1=2t$.
All error bars here indicate one standard deviation from multiple measurements.}
\end{figure*}

To experimentally simulate the non-unitary dynamics driven by Hamiltonian (\ref{eq1}), we embed the dissipative Hamiltonian (\ref{eq1}) into a larger Hilbert space composed of atomic momentum states. As illustrated in Fig.~\ref{fig2}(a), a momentum lattice is generated by imposing multiple pairs of counter-propagating, far-detuned Bragg lasers (with the wavelength $\lambda_0=1064nm$) on a Bose-Einstein condensate (BEC) of $\sim 10^5$ $^{87}$Rb atoms in a weak optical dipole trap \cite{suppinfo,Gadway2015}. The frequencies of the Bragg lasers are carefully designed to couple $8$ discrete momentum states $p_n=2n\hbar k$ ($k=2\pi/\lambda_0$ and $n=0,1,...7$), which form a synthetic lattice of finite size, with individually tunable Bragg-assisted tunneling strength $t_n$ between adjacent sites $|n-1\rangle$ and $|n\rangle$; see Figs.~\ref{fig2}(b) and (c). A unitary kick is then introduced through a square-wave modulation $t_2=t_z(\tau)$ for the inter-site coupling $|1\rangle\leftrightarrow |2\rangle$. Consistent with Eq.~(\ref{eq2}), $t_z(\tau)=t_0$ for $jT\le \tau < jT+\tau_0$, while vanishes for other time intervals. Treating momentum-lattice sites $|n\ge 2\rangle$ as a reservoir, we find that dynamics within the two-dimensional subspace spanned by $\{|0\rangle,|1\rangle\}$ to be dissipative, and effectively driven by Hamiltonian (\ref{eq1}) with $\gamma_0 \sim t_0^2/t$~\cite{Lapp2019,Gou2020}. As such, we implement an effectively dissipative two-level system in momentum space, whose dissipation originates from unitary kicks that, with kick frequency $\Omega$ and intensity $\gamma_0$, periodically couple the system with a reservoir.

We study both the PT symmetry breaking transition and the quantum Zeno (anti-Zeno) effects through the dissipative dynamics. Specifically, we initialize the atoms in the state $|0\rangle$, and let them evolve for a short time $\tau_e$, before applying a time-of-flight image to record the atomic probability distribution $P_n$ for each momentum lattice site, normalized by the total atom population over the momentum lattice~\cite{suppinfo}. We typically let $\tau_e$ be two or three modulation periods, limited by both the finite size of the reservoir and the decoherence time of the system~\cite{Gou2020, Xie2020}.
Under the passive PT symmetric Hamiltonian (\ref{eq1}), the PTS and PTB phases can be dynamically differentiated by the corrected probability
\begin{align}
\mathcal{P}^c_s =e^{\gamma_0 \tau_0 \frac{\tau_e}{T}}(P_0+P_1),
\end{align}
which reflects the time evolution of the squared state norm within the synthetic subspace driven by the Hamiltonian $H_{PT}$. It follows that $\mathcal{P}^c_s$ should be on the order of unity in the PTS phase, while it should exponentially grow with time in the PTB phase.
To further characterize quantum Zeno and anti-Zeno regimes, we probe the averaged effective loss rate $\bar{\gamma}$
\begin{equation}\label{eq3}
 e^{-2\bar{\gamma}\tau_e} = 1-\mathcal{P}_r,
\end{equation}
where $\mathcal{P}_r$ is the total loss rate from the dissipative two-level system, with $\mathcal{P}_r=\sum_{n\ge 2} P_n$.

In Fig.~\ref{fig3}(a), we show the measured corrected probability $\mathcal{P}_s^c$ and the effective loss rate $\bar{\gamma}$ across the PT phase transition $\mathcal{M}_0\to\mathcal{L}_0$ at a high kick frequency $\Omega/t=10$ and with increasing kick intensity $\gamma_0$.
The measured corrected probability (middle panel) becomes exponentially large beyond the exceptional point at $\gamma_0/t\sim 2$ (dash-dotted vertical line from upper panel).
In the PTS (PTB) phase $\mathcal{M}_0$ ($\mathcal{L}_0$), the effective loss rate of the synthetic two-level system increases (decreases) with increasing $\gamma_0$ (see lower panel), indicating quantum anti-Zeno (Zeno) regime. The effective loss rate $\bar{\gamma}$ peaks near the exceptional point, consistent with the theoretical predication that the quantum Zeno to anti-Zeno transition should coincide with the PTB-PTS transition.

\begin{figure}[]
\includegraphics[width=0.4\textwidth]{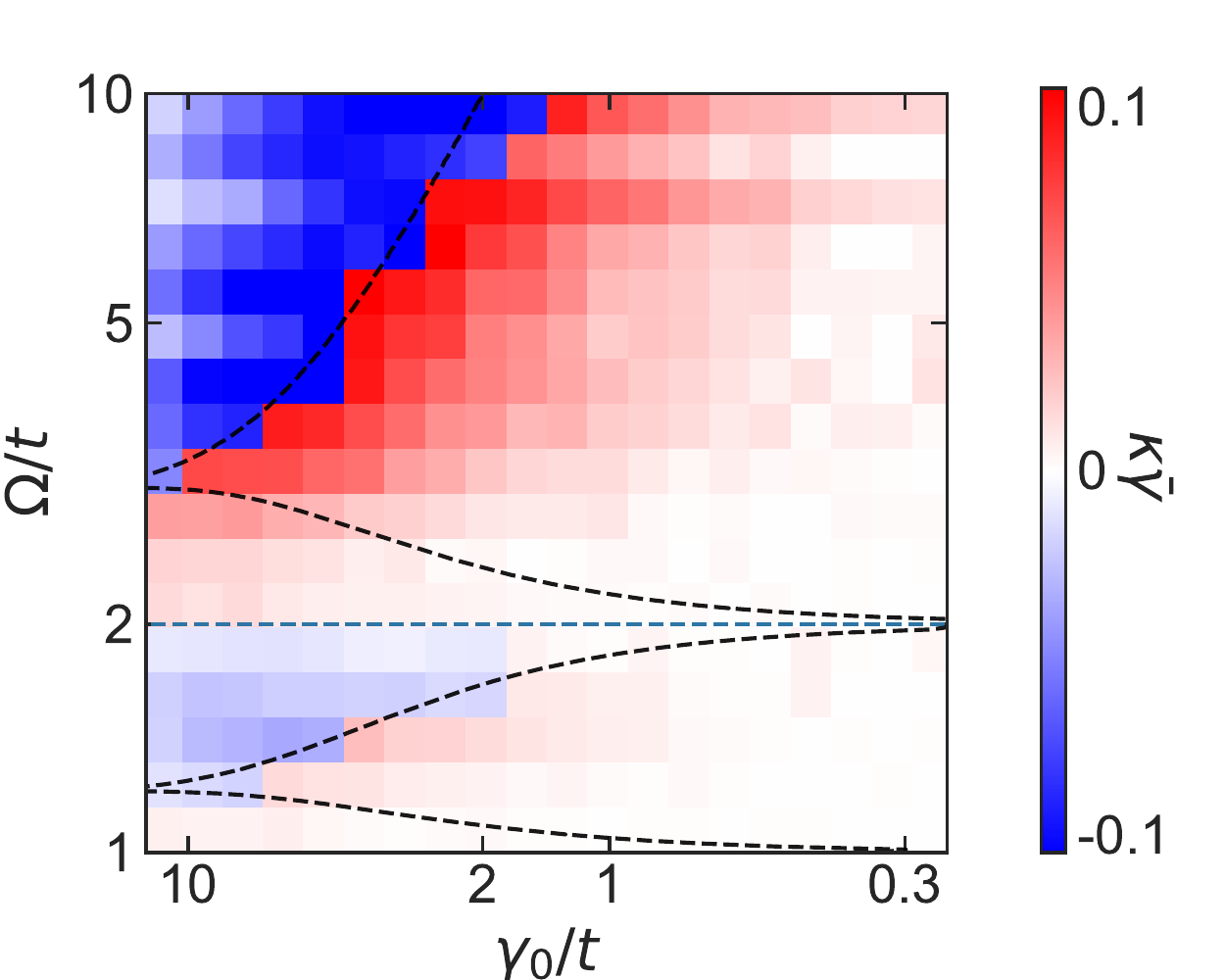}
\caption{\label{fig4} (Color Online)
Phase diagram for quantum anti-Zeno to Zeno transition from experimental data. Color contour is the measured $\kappa\bar{\gamma}$ (see main text for definition).
The black dashed lines indicate the exceptional points [see Fig.\ref{fig1}(c)], and the blue dashed line corresponds to the critical kick frequency $\Omega_1=2t$.
The blue (red) regions correspond to quantum Zeno (anti-Zeno) regimes.
All $\bar{\gamma}$ are measured after the system evolves for two modulation periods, while we set $t=2\pi\times 1.03(2)~{\rm kHz}$ and $\tau_0t/(2\pi)=0.1$ for all measurements. The measured critical anti-Zeno to Zeno transition points are consistent with the exceptional points between the $\mathcal{M}_j$ and $\mathcal{L}_j$ regions.
}
\end{figure}

However, such is not the case at lower kick frequencies. As illustrated in Fig.~\ref{fig3}(b), when $\gamma_0$ is tuned at a fixed $\Omega_/t=2.5$, $\bar{\gamma}$ increases monotonically across the transition $\mathcal{M}_0\to\mathcal{V}_1$ at $\gamma_0/t=2.8$ (dash-dotted line), suggesting both the PT symmetric $\mathcal{M}_0$ and the PT broken $\mathcal{V}_1$ belong to the quantum anti-Zeno regime. Note that at the critical kick frequencies, for instance $\Omega_1=2t$, $\bar{\gamma}$ also increases with increasing $\gamma_0$. Thus quantum anti-Zeno effects survive at the boundaries between $\mathcal{V}_j$ and $\mathcal{L}_j$ in the PTB phase.

Apart from tuning $\gamma_0$, both the PT-symmetry breaking transition and quantum Zeno to anti-Zeno transition can be crossed by changing the kick frequency, which amounts to traversing the phase diagram Fig.~\ref{fig1} vertically.
Figure~\ref{fig3}(c) shows the measured $\bar{\gamma}$ across multiple PT phase transitions by increasing $\Omega$ with a fixed $\gamma_0/t\sim 9$. The measured effective loss rate $\bar\gamma$ peaks near the PT-symmetry phase boundary between $\mathcal{M}_j$ and $\mathcal{L}_j$, consistent with the coincidence of the two transitions according to the theoretical phase diagram. Further, a local minimum in $\bar\gamma$ is found near the critical kick frequency $\Omega_1=2t$ (lower panel), where a ``slow mode'', i.e., the eigenstate with the smaller imaginary eigenvalue, dominates the dynamics~\cite{Li2020}.
We emphasize that, the occurrence of quantum anti-Zeno effect in the PTB regime is unique to slow modulations. For fast modulations ($\Omega/t \gg 1$, where the transition $\mathcal{M}_0\to\mathcal{L}_0$ lies), increasing the kick rate is similar to enlarging the dissipation rate in the continuous case~\cite{Facchi2008, Li2020}. There, only a single transition point from the quantum anti-Zeno to Zeno regime exists, which occurs exactly at the exceptional point.

While the experimental measurements in Fig.~{\ref{fig3} qualitatively agree with theoretical predications, quantitative deviations
exist, which mainly derive from two sources.
First, the kick intensity $\gamma_0$ in the effective Hamiltonian (\ref{eq1}) would deviate from the perturbative expression $\gamma_0 \sim t_0^2/t$ when either the coupling $t_0$ or the evolution time become sufficiently large.
This is the main reason for the slight discrepancy between the location of the maximum loss rate in Fig.~{\ref{fig3}(a), either numerically simulated (dashed and solid lines) or experimentally measured, and that of the theoretically predicted exceptional point using the perturbative kick intensity (dash-dotted).
Second, high-order, non-resonant coupling terms play an important role in our experiment, as is manifest in Fig.~\ref{fig3} where the experimental data agree better with simulations considering the non-resonant coupling terms (solid lines). As non-resonant couplings enable the $|0\rangle\leftrightarrow |-1\rangle$ transition, the population of the $|-1\rangle$ state leads to an underestimation of loss for a finite evolution time. Other factors, for example, interaction-induced self-trapping in the momentum lattice and the momentum broadening due to the weak trap potential~\cite{Gou2020, An2018, Xie2020}, also lead to underestimations of the loss rate.

Finally, we map out the phase diagram for quantum Zeno to anti-Zeno transition by sweeping $t_0$ (hence $\gamma_0$) for a set of fixed $\Omega$, and plotting the quantity $\kappa\bar{\gamma}$ with $\kappa = {\rm sgn}(\Delta\bar{\gamma}/\Delta\gamma_0)$; see Fig.~\ref{fig4}. Here the difference $\Delta\bar{\gamma}/\Delta\gamma_0$ is calculated from experimental data for each fixed $\Omega$. By definition,
$\kappa\bar{\gamma}<0$ ($\kappa\bar{\gamma}>0$) represents the quantum Zeno (anti-Zeno) regime. At the lower-right corner of Fig.~\ref{fig4}, $\kappa\bar{\gamma}$ is close to zero, due to a vanishing $t_z$ and a disconnected reservoir. At the upper-left corner, $\kappa\bar{\gamma}$ also approaches zero, as loss to the reservoir is suppressed, which is equivalent to the standard quantum Zeno effect in the case of continuous, strong couplings.
Most importantly, by superimposing the boundaries of PT transitions (black dashed) and the critical kick frequency (blue dashed), it is clear that our measured phase diagram in Fig.~\ref{fig4} agrees well with the theoretical prediction in Fig.~\ref{fig1}(c), thus confirming the following correspondence
\begin{eqnarray}
 \mathcal{V}_j{\rm (PTB)}, \mathcal{M}_j{\rm (PTS)} &\leftrightarrow & \text{anti-Zeno}, \nonumber\\
 \mathcal{L}_j{\rm (PTB)} &\leftrightarrow & \text{Zeno}.
\end{eqnarray}
Such a relationship reveals the deep connection between PT transition and quantum Zeno effects.

To conclude, we have experimentally established the connection between the quantum Zeno effect and PT phases in a dissipative Floquet system:
while the PTS phase generally leads to the quantum anti-Zeno effect, both quantum Zeno and anti-Zeno effects can occur in the PTB region.
Crucially, the quantum-Zeno regimes are bounded by a discrete set of critical coupling frequencies, and by exceptional points. Besides
shedding new lights on the relation of quantum measurements and dynamics of non-Hermitian systems, our experiment also offers a new way of simulating PT physics using cold atoms, which is readily extendable to higher dimensions~\cite{suppinfo}.
While quantum Zeno effects and the associated quantum Zeno subspace~\cite{Facchi2008} generally exist for multi-level systems, the scalability of the correspondence considered here to higher dimensions is an interesting open question which we leave to future studies.

Further, the above analyses are all within the scope of single-particle physics, without considering the effect of interactions. Specifically, many-body interactions in the momentum lattice assume the form of density-dependent, attractive on-site potentials~\cite{An2018}. When the atomic density or the scattering length is large enough, atoms in momentum space exhibit the so-called interaction-induced localization~\cite{An2018, Xie2020}. Since the quantum Zeno dynamics can also be regarded as a form of localization (or stabilization) within the quantum Zeno subspace~\cite{Barontini2013,Patil2015,Peise2015}, it will be interesting to study the interplay between interactions and quantum Zeno effects in future experiments~\cite{Everest2017,Rubio2019,Bouganne2020,Choi2020}.

We acknowledge support from the National Key R$\&$D Program of China under Grant Nos. 2018YFA0307200, 2016YFA0301700 and 2017YFA0304100, the National Natural Science Foundation of China under Grant Nos. 91636104, 11974331 and 91736209, the Natural Science Foundation of Zhejiang province under Grant No. LZ18A040001, and the Fundamental Research Funds for the Central Universities.

\bibliographystyle{apsrev4-1}

\bibliography{PTvsqze}

\clearpage
\begin{widetext}
\newcommand{\smtitle}[1]{\begin{center}\large{\textbf{#1}}\end{center}\vskip 2em}
\renewcommand{\thefigure}{S\arabic{figure}}
\renewcommand{\thetable}{S\Roman{table}}
\setcounter{figure}{0}
\renewcommand{\theequation}{S\arabic{equation}}
\setcounter{equation}{0}

\smtitle{Supplemental Materials}

Here we provide more details on the experimental procedure, the theoretical simulation, and the effective Hamiltonian of the system.

\subsection{Experimental settings}

The $^{87}$Rb BEC is prepared in an optical dipole trap by evaporative cooling for $\sim 18~\text{s}$. The multiple discrete momentum states are coupled with multi-frequency Bragg laser pairs. The different frequency components are imprinted by two acoustic optical modulators. One shifts the frequency of the incoming beam by $-100~{\rm MHz}$, and another shifts it by $100~{\rm MHz}-\sum_n \nu_n/2\pi$ ($n\ge 1$) with $\nu_n = 4(2n-1)\hbar k^2/2m$ (see the main text). As a consequence, the transition between the two momentum states, $|n-1\rangle\leftrightarrow |n\rangle$, can be resonantly triggered by the $\{\omega_+,\omega_n\}$ laser pair.

After the system evolves for a finite time $\tau_e$, we directly resolve the populations in each momentum state by letting the atoms fall freely in space for 20 ms with all lasers switched off, before the atoms are imaged by a CCD camera. Atoms with different momenta get separated in the $x$-direction along which the Bragg beams are applied (see Fig.~1 in the main text). To obtain the relative populations in each state, we integrate the image in the $y$-direction, and then fit the data with a $10$-peak Gauss function, $\mathcal{A}(x)=\sum_{n=-1}^{8} A_n \text{exp}\left[-\left(\frac{x-nd}{a}\right)^2\right]$. Normalizing the resulting amplitude $\mathcal{A}_n$ by $\sum_{n=-1}^{8} \mathcal{A}_n$, we finally get the atom probability distribution on each site, $P_n$.

\subsection{Effective Hamiltonian with off-resonant terms}

Following the theory of light-atom interaction in Ref.~\cite{Gadway2015}, we obtain the effective time-dependent full Hamiltonian
\begin{equation}\label{eqs2}
 H_{\rm eff} = \sum_n\sum_i\hbar\frac{\Omega_+\Omega_i}{4|\Delta|}e^{i[(\omega_+-\omega_i-4(2n-1)\hbar k^2/2m)t+(\phi_+-\phi_i)]}|n-1\rangle\langle n| + {\rm H.c.}
\end{equation}
with $\phi_+$ and $\phi_i$ the phases of the $\omega_+$ beam and $\omega_i$ component respectively (see Fig.~1 in the main text). We simply let $\phi_+=0$, and $\phi_i$ be the modulated phase relative to $\phi_+$ from the AOM. As we choose $\omega_i = \omega_+ - 4(2i-1)\hbar k^2/2m$, the simplified ideal model can be obtained by considering only the resonant terms, i.e., letting $i=n$, as
\begin{equation}\label{eqs1}
 H^{(0)} = \sum_n \hbar t_n |n-1\rangle \langle n| + {\rm H.c.}
\end{equation}
with $t_n = e^{-i\phi_n}\Omega_+\Omega_n/4|\Delta|$. This gives the general tight-binding form for a momentum-state chain. If we simply treat the $n \ge 2$ part as an effective reservoir, and apply the second-order perturbation with $t_2=t_z$ and $t_{n\neq 2}=t$, the loss rate of site $|1\rangle$ should approximately be $\gamma_0 \sim t_z^2/t$ \cite{Lapp2019}. Then we obtain the dissipative two-level Hamiltonian in the main text.

\begin{figure*}[]
\includegraphics[width=0.5\textwidth]{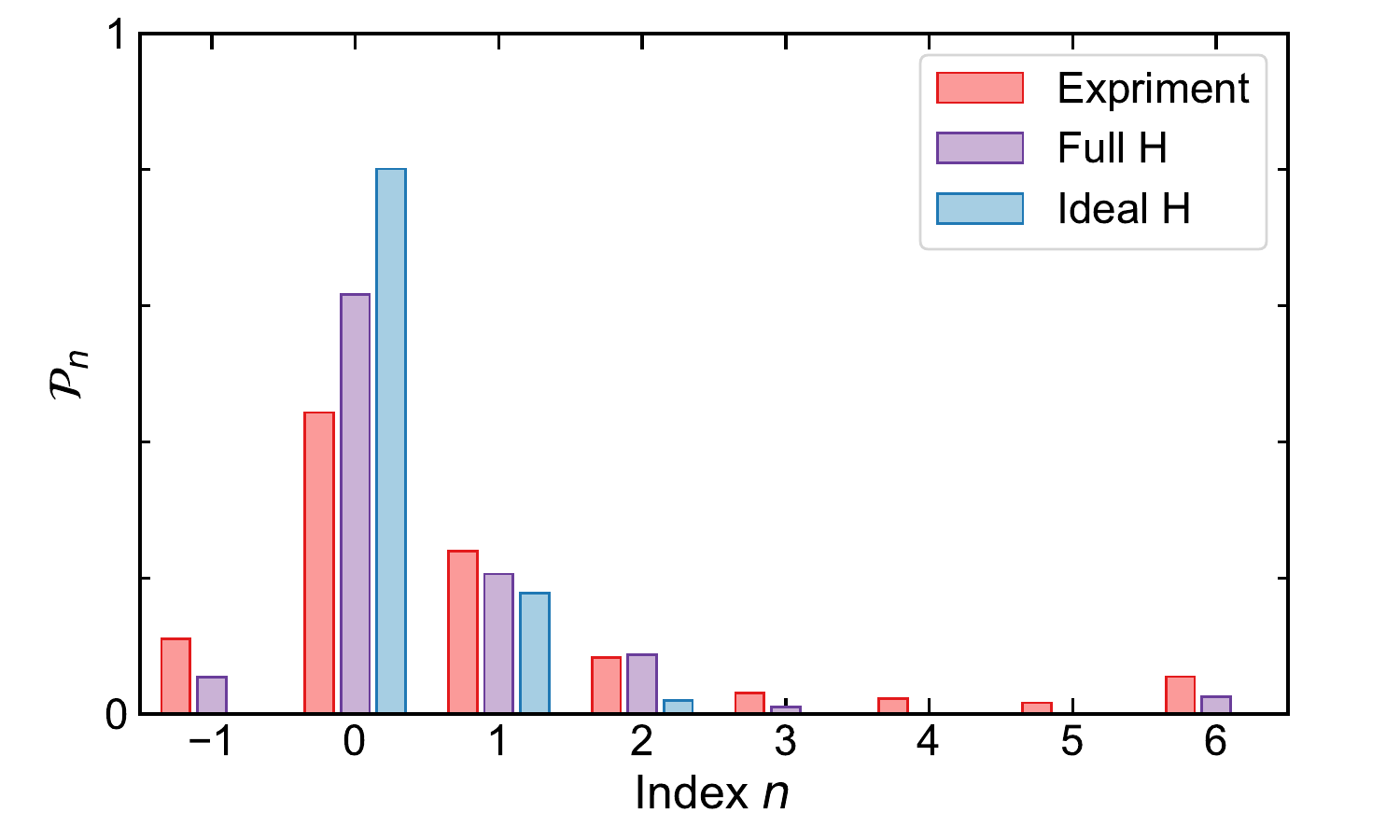}
\caption{Population distribution after a $6$-state chain evolves for $0.5$ms. A comparison between the experimental results and those from numerical simulations shows the effect of the non-resonant Bragg diffractions. Simulations using the ideal Hamiltonian (\ref{eqs1}) and the full Hamiltonian adopt the same coupling parameters $t_n=2\pi\times 1.50(2)~{\rm kHz}$.}\label{figS1}
\end{figure*}

Then, clearly, the $\ell$th-order non-resonant terms, responsible for the transition $|n-1\rangle\leftrightarrow|n\rangle$, can be induced by $\{\omega_+, \omega_{n-\ell}\}$ and $\{\omega_+, \omega_{n+\ell}\}$ laser pairs with detunings of $\mp 8\ell\hbar k^2/2m$ respectively. These terms are given by
\begin{equation}
 H^{(\ell)} = \sum_n \hbar t_{n\pm \ell} e^{\pm i8\ell(\hbar k^2/2m)t} |n-1\rangle\langle n| + \text{H.c.},
\end{equation}
leading the full Hamiltonian $H_{\rm eff}=\sum_\ell H^{(\ell)}$. In our experiment, $8\hbar k^2/2m$ corresponds to $\sim 2\pi\times 16.2~\text{kHz}$.

To demonstrate the effect of non-resonant terms, we simulate a $0.5$ms time evolution with the time-dependent effective full Hamiltonian (\ref{eqs2}), and make a comparison to the experimentally measured populations, and to simulation results from the effective ideal Hamiltonian (\ref{eqs1}). As shown in Fig.~\ref{figS1}, the full Hamiltonian describes our system better, and the states $|-1\rangle$ and $|6\rangle$ indeed get populated due to the non-resonant couplings, even when they are not resonantly coupled.

\subsection{Quantum Zeno subspace}

\begin{figure}[]
\includegraphics[width=0.5\textwidth]{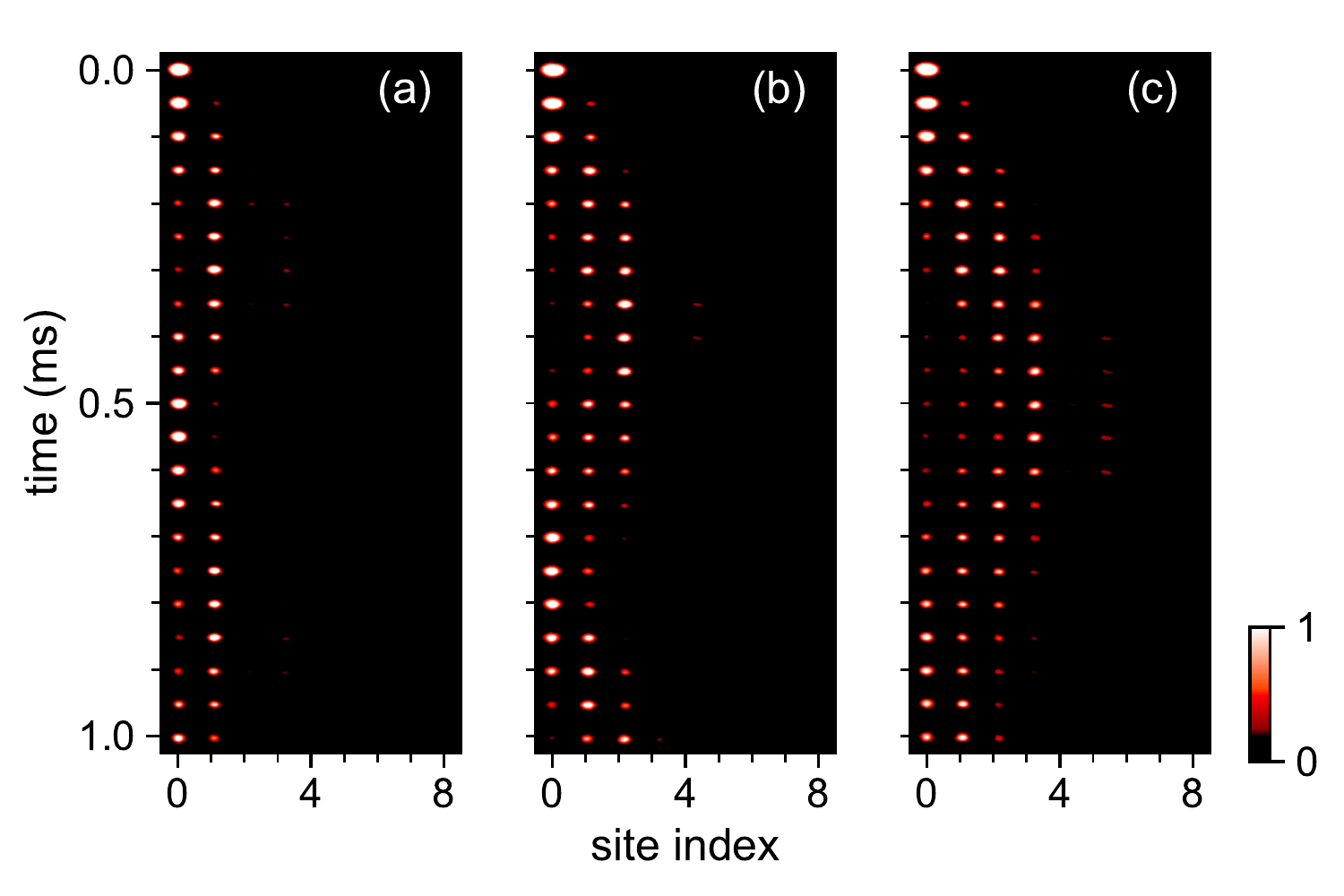}
\caption{\label{figS2}
Tailoring the Hilbert space with quantum Zeno effects. We work with the $8$-site momentum lattice illustrated in Fig.~\ref{fig2}(c) with the tunneling strength $t=2\pi\times 1.03(2)~{\rm kHz}$ except at locations of strong Zeno couplings: (a) $t_3=3t$, (b) $t_4=3t$ and (c) $t_5=3t$. The measured time evolutions of $P_n$ are confined within the left-hand subspaces, with small leakage to the outside due to finite Zeno coupling strengths. Here we fix the kick frequency $\Omega/t=10$, and $\tau_0/T = 1$ for all cases.
}
\end{figure}

In this section, we demonstrate the extendability of our experimental configuration by observing quantum Zeno effects withs an enlarged dissipative subspace. At the upper-left corner of the phase diagram in Fig.~\ref{fig4}, $\bar{\gamma}$ approaches zero, as loss to the reservoir is suppressed due to the quantum Zeno effect.
In this deep PTB regime, the Hilbert space is tailored to the quantum Zeno subspace and the system can be described by an approximately Hermitian Hamiltonian~\cite{Facchi2008}. Building upon this understanding, in Fig.~\ref{figS2}, we show a series of examples of tailoring the quantum Zeno subspace by enlarging the dissipative subspace.
Specifically, when we shift the Zeno coupling $t_z$ to $|2\rangle\leftrightarrow |3\rangle$, the effective dissipation rate for site $|2\rangle$ in Fig.~\ref{figS2}(a) should be $\gamma_0\sim 9t$. Now the oscillations take place only in a subspace $\mathcal{H}_z$ spanned by $\{|0\rangle, |1\rangle\}$, and the site $|2\rangle$ becomes a boundary as the quantum Zeno effect guarantees it maintaining its initial condition, $P_2=0$. The imperfect Rabi oscillation originate from several sources, for instance, finite strength of $\gamma_0$, high-order non-resonant coupling terms, and many-body interactions. The size of the Zeno subspace can be easily tuned by shifting the position of the strong Zeno coupling. As shown in Fig.~\ref{figS2}(b), we let $t_{4}/t=3$ and find that, the time evolutions are confined within the subspace $\mathcal{H}_z=\{|0\rangle, |1\rangle, |2\rangle\}$. Furthermore, the subspace gets enlarged with $|3\rangle$ included once we shift Zeno coupling to $|4\rangle\leftrightarrow |5\rangle$; see Fig.~\ref{figS2}(c).
These experiments suggest that, combined with the flexible control offered by momentum lattice,
quantum Zeno effects can be used to map the whole ensemble into multiple Zeno subspaces with arbitrary sizes.
The approximately coherent dynamics within the Zeno subspace provide opportunities to explore a variety of non-classical phenomena, for example, the generation of Sch\"odinger cat states~\cite{Barontini2015}. We also note that, while the multilevel systems studied in this section do not possess PT symmetry, the scalability of our system as well as the demonstration of quantum Zeno subspaces pave the way for future studies of the correspondence between PT symmetry and quantum Zeno effects in higher dimensions.

\clearpage
\end{widetext}
\clearpage
\end{document}